\documentclass[final,5p,times,twocolumn]{elsarticle}
\usepackage{amsmath,amssymb,amsfonts}
\usepackage{graphicx}
\usepackage{slashed}
\usepackage{citesort}

\usepackage{dsfont}
\allowdisplaybreaks[1]
\newcommand{\beq}{\begin{equation}}
\newcommand{\eeq}{\end{equation}}

\newcommand{\Lagr}{\mathcal{L}}

\newcommand{\reff}[1]{(\ref{#1})}
\begin{document}

\begin{frontmatter}
\title{\hfill{\scriptsize HISKP--TH--10/28, FZJ-IKP-TH-2010-25}\\ 
Chiral dynamics of the $S_{11}(1535)$ and $S_{11}(1650)$ resonances revisited}

\author[Regensburg]{Peter~C.~Bruns}
\author[Bonn]{Maxim~Mai}
\author[Bonn,Julich]{Ulf-G.~Mei{\ss}ner}

\address[Regensburg]{Institut f\"ur Theoretische Physik, Universit\"at Regensburg,
             D-93040 Regensburg, Germany}
\address[Bonn]{Helmholtz--Institut f\"ur Strahlen- und Kernphysik (Theorie) 
   and Bethe Center for Theoretical Physics, Universit\"at Bonn, D-53115 Bonn, Germany}
\address[Julich]{Institut f\"ur Kernphysik, Institute for Advanced Simulation, 
   and J\"ulich Center for Hadron Physics, Forschungszentrum J\"ulich, D-52425  J\"ulich, Germany}
	\begin{abstract}
	We analyze s-wave pion-nucleon scattering in a unitarized chiral effective
	 Lagrangian including all dimension two contact terms. We find that both the
	 $S_{11}(1535)$ and the $S_{11}(1650)$ are dynamically generated, but the
	 $S_{31}(1620)$ is not. We further discuss the structure of these dynamically
	 generated resonances.
	\end{abstract}

\end{frontmatter}

		\section{Introduction}

Pion-nucleon scattering has traditionally been the premier reaction to study the
 resonance excitations of the nucleon. In particular, in the $S_{11}$ partial wave,
 one finds two close-by resonances at 1535 and 1650~MeV, which overlap within their
 widths of about 100~MeV. It was pointed out early in the framework of unitarized
 coupled-channel  chiral perturbation theory \cite{Kaiser:1995cy} that this resonance
 might not be a three-quark (pre-existing) resonance but rather is generated by
 strong channel couplings, with a dominant $K\Sigma - K\Lambda$ component in its
 wave function.  This analysis was extended in Ref.~\cite{Inoue:2001ip}, where
 within certain approximations the effects of 3-body $\pi\pi N$ channels were
 also included. Further progress was made in  Ref.~\cite{Nieves:2001wt}, where
 the $S_{11}$ phase shift was fitted from threshold to about $\sqrt{s} \simeq 2\,$GeV
 together with cross section data for $\pi^- p\to \eta n$ and $\pi^- p\to K^0 \Lambda$
 in the respective threshold regions. This led to a satisfactory description of
 the $S_{11}$ phase and a reasonable description of the inelasticity up to  the
 $\eta N$ threshold. Two poles were found corresponding to the  $S_{11}(1535)$
 and the $S_{11}(1650)$ resonances together with a close-by unphysical  pole on
 the first Riemann sheet. More recently, it was pointed out in a state-of-the-art
 unitary meson-exchange model that there is indeed  strong resonance interference
 between the two $S_{11}$ resonances, as each of these resonances provides an
 energy-dependent background in the region of the other \cite{Doring:2009yv}.

In view of these developments and our attempts to construct a unitary and gauge-invariant
 model for Goldstone-boson photoproduction off nucleons based on coupled-channel
 unitarized chiral perturbation theory~\cite{Borasoy:2007ku}, we consider in this
 letter the two s-waves $S_{11}$ and $S_{31}$ in pion-nucleon scattering. We work in
 the framework of a coupled-channel Bethe-Salpeter equation (BSE)  including in the
 driving potential all local terms of second order in the chiral counting, thus going
 beyond the often used approximation of simply including the leading order Weinberg-Tomozawa
 interaction. Further, we do  not perform the often used on-shell approximation. Note
 that $K^-p$ scattering including such dimension two terms was already analyzed in a
 framework equivalent to the on-shell approximation of the Bethe-Salpeter equation in
 Refs.~\cite{Borasoy:2005ie,Oller:2006jw,Borasoy:2006sr}.      Our investigation is
 restricted to center-of-mass energies below 1.8~GeV, as required for the future meson
 photoproduction studies. As we will show, both resonances in the $S_{11}$ partial
 wave are dynamically generated, even if the scattering data are fitted only up to
 $\sqrt{s} = 1.56\,$GeV. Quite in contrast, the $S_{31} (1620)$ resonance is not
 generated by the coupled-channel dynamics. We also analyze the structure of
 the dynamically generated resonances as revealed through their coupling to the various
 meson-baryon channels.

		\section{Formalism}

We consider the process of meson--baryon scattering at low energies. 
 The s-wave interaction near the thresholds is dominated by the 
 Weinberg-Tomozawa contact term, derived from the effective chiral Lagrangian
	\begin{align}
	\Lagr^{(1)}_{\phi B}&=\langle \bar{B} (i\gamma_\mu D^\mu-m_0)B\rangle
	+\frac{D/F}{2}\langle \bar{B}\gamma_\mu \gamma_5[u^\mu,B]_\pm \rangle ~,
	\end{align}
 where $\langle\ldots\rangle$ denotes the trace in flavor space, 
 $D_\mu B :=\partial_\mu B +\frac{1}{2}[[u^\dagger,\partial_\mu u],B]$, 
 $m_0$ is the baryon octet mass in the chiral SU(3) limit, 
 and $D$, $F$ are the axial coupling constants. The relevant degrees 
 of freedom are the Goldstone bosons described by the traceless 
 meson matrix $U$,
	\begin{equation}\label{eqn:fields}
	U =\exp\Bigl(i\frac{\phi}{F_0}\Bigr)\,, ~ \phi=\sqrt{2}\begin{pmatrix}
	\frac{\pi^0}{\sqrt{2}}+\frac{\eta}{\sqrt{6}} \!\!&\!\! \pi^+ \!\!&\!\! K^+ \\
	\pi^- \!\!&\!\! -\frac{\pi^0}{\sqrt{2}}+\frac{\eta}{\sqrt{6}} \!\!&\!\! K^0 \\
	K^- \!\!&\!\! \bar{K}^0 \!\!&\!\! -\frac{2}{\sqrt{6}}\eta
	\end{pmatrix}~,
	\end{equation}
 where $F_0$ is the meson decay constant in the chiral limit, 
 and the low-lying baryons are collected in a traceless matrix
	\begin{equation}\label{eqn:baryonmatrix}
	B=\begin{pmatrix}
	\frac{\Sigma^0}{\sqrt{2}}+\frac{\Lambda}{\sqrt{6}} & \Sigma^+ & p \\
	\Sigma^- & -\frac{\Sigma^0}{\sqrt{2}}+\frac{\Lambda}{\sqrt{6}}& n \\
	\Xi^-& {\Xi}^0 & -\frac{2}{\sqrt{6}}\Lambda
	\end{pmatrix}~.
	\end{equation}
 We set external currents to zero except for the scalar one, 
 which is set equal to the quark mass matrix, $s=\mathcal{M}:=\textrm{diag}(m_u, m_d, m_s)$.  We furthermore use
	\begin{align}
	u^2:=U  ~,\quad
	u^\mu:=iu^{\dagger}\partial^\mu u - iu\partial^\mu u^{\dagger} ~,\nonumber\\
	\quad
	\chi_\pm:=u^{\dagger}\chi u^{\dagger}\pm u\chi^{\dagger}u ~,\quad
	\chi:=2B_0\, s ~,
	\end{align}
 where the constant $B_0$ is related to the quark condensate in the chiral limit.

The Weinberg-Tomozawa contact term mentioned above stems from the covariant 
 derivative $D_\mu B\,$, and is of first order in the chiral power counting. 
 Most chiral unitary approaches restrict their meson-baryon potential to 
 this interaction, which generates the leading contribution to the s-wave 
 scattering lengths. This approach has been remarkably successful in many
 cases, see 
 e.~g.~\cite{Kaiser:1995eg,Kaiser:1995cy,Oset:1997it,Oller:2000ma,Oller:2000fj,Lutz:2001yb}. 
 However, at first chiral order, there are also the Born graphs, describing 
 the $s$-channel and $u$-channel exchanges  of an intermediate nucleon. The full 
 inclusion of these graphs in the driving term of the Bethe-Salpeter equation 
 leads to conceptional and practical difficulties, which have not yet been
 solved to the best of our know\-ledge:
 (i) Iteration of the $s$-channel exchange Born graphs will generate various 
 contributions leading to a renormalization of the various baryon masses 
(and wave function renormalizations), which are usually set to their 
 physical values in the loop functions of the chiral unitary approach. 
 These contributions would thus have to be dropped. In view of a later 
 application to photoproduction, such a non-perturbative treatment of 
 $s$-channel exchanges leads to complications with gauge invariance because 
 the self-energies are linked (via a Ward-Takahashi identity) to the 
 electromagnetic baryon form factors, which would also have to be treated 
 in a corresponding (non-perturbative) fashion.
 (ii) Iteration of the $u$-channel diagram, on the other hand, leads to all kinds 
 of genuine multi-loop topologies, as there is no factorization into simple 
 one-loop terms any more. The corresponding integral equation could only 
 be solved numerically, e.g. by employing a Wick rotation and a four-momentum
 cutoff. Problems with gauge invariance would also occur here. In the
 literature, the $u$-channel Born diagrams were usually treated within some 
 approximation which effectively reduced the solution of the BSE to products 
 of one-loop terms, or included perturbatively to guarantee a matching to 
 ChPT amplitudes up to a given order, see e.g.~\cite{Meissner:1999vr}.
 All these approximations, however, destroy the exact correspondence of the 
 individual terms in the solution of the BSE to dimensionally regularized
 Feynman  graphs, which is crucial in our approach to photoproduction.
 Therefore, we will approximate our interaction kernel by a sum of contact
 terms. To go beyond the simple Weinberg-Tomozawa potential, we shall 
 include the full set of meson-baryon vertices from the second order 
 chiral Lagrangian. These terms may lead to sizeable corrections to the 
 leading-order results, see e.~g. the calculation of NNLO corrections 
 on meson-baryon scattering lengths within SU(3) ChPT~\cite{Mai:2009ce}.
 The pertinent Lagrangian density was first constructed in
 \cite{Krause:1990xc} and reads in its minimal form~\cite{Frink:2004ic}
	\begin{align}\label{eqn:LAGR}
	&\Lagr^{(2)}_{\phi B}= b_{D/F} \langle\overline B
	\big[\chi_+,B\big]_\pm\rangle
	+b_0 \langle\overline B B\rangle \langle\chi_+\rangle\nonumber\\
	&+b_{1/2} \langle\overline B  \Big[u_\mu,\big[u^\mu,B\big]_\mp\Big]\rangle
	+b_3 \langle\overline B \Big\{ u_\mu,\big\{ u^\mu,B\big\}\Big\}\rangle
	+b_4 \langle\overline B  B\rangle \langle u_\mu u^\mu\rangle \nonumber\\
	&+ib_{5/6}  \langle\overline B\sigma^{\mu\nu} \Big[\big[u_\mu,u_\nu\big], B\Big]_\mp\rangle
	+ib_7 \langle\overline B\sigma^{\mu\nu} u_\mu\rangle  \langle u_\nu B\rangle \nonumber\\
	&+ \frac{i\,b_{8/9}}{2m_0}\Big( \langle\overline B \gamma^\mu\Big[u_\mu,\big[u_\nu,\big[D^\nu, B\big]\big]_\mp\Big]\rangle+\langle\overline B \gamma^\mu\Big[D_\nu,\big[u^\nu,
	\big[u_\mu,B\big]\big]_\mp\Big]\rangle\Big) \nonumber\\
	&+\frac{i\,b_{10}}{2m_0}\Big( \langle\overline B 
	\gamma^\mu\Big\{ u_\mu,\big\{ u_\nu,\big[D^\nu,B\big]\big\}\Big\}\rangle+\langle\overline B\gamma^\mu\Big[D_\nu,\big\{ u^\nu,
	\big\{ u_\mu,B\big\}\big\}\Big]\rangle\Big) \nonumber\\
	&+\frac{i\,b_{11}}{2m_0}\Big( 2\langle\overline B \gamma^\mu 
	\big[D_\nu,B\big]\rangle \langle u_\mu u^\nu\rangle\nonumber\\
	&\qquad\qquad\quad +\langle\overline B \gamma^\mu B\rangle 
	\langle\big[D_\nu,u_\mu\big]u^\nu + u_\mu \big[D_\nu,u^\nu\big]\rangle   \Big)~,
	\end{align}
 with the $b_i$ the pertinent dimension-two low energy constants (LECs).
 The LECs $b_{0,D,F}$ are the so-called {\it symmetry breakers} while the
 $b_i$ $(i= 1,\ldots, 11)$ are referred to as {\it dynamical} LECs.

The strict perturbative chiral expansion is only applicable at low energies.
 Moreover, it certainly fails in the vicinity of resonances. The purpose of
 the present work is the extension of the range of applicability of the
 low-energy effective theory by means of a coupled channel Bethe-Salpeter
 equation (BSE). Introduced in~\cite{Salpeter:1951sz} it has been proven to
 be very useful both in the purely mesonic and in the meson-baryon
 sector~\cite{Kaiser:1995eg,Kaiser:1995cy,Oset:1997it,Oller:2000ma,Oller:2000fj,Lutz:2001yb}.
 In contrast to perturbative calculations this approach implements two-body unitarity
 exactly and in principle allows to generate resonances dynamically. Due to the exact
 correspondence of the Bethe-Salpeter scattering amplitude with an infinite sum of
 dimensionally regularized Feynman graphs, we can use our solution of the BSE as
 an extended vertex in a model amplitude for meson photoproduction and arrive at
 a natural and straightforward way to implement gauge invariance in a chiral
 unitary framework (for details on the construction principles,
 see~\cite{Borasoy:2007ku}).

\begin{figure}[t]
	\begin{center}
	\includegraphics[width=1.0\linewidth]{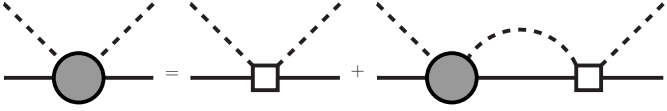}
	\end{center}
	\caption{Symbolical representation of the Bethe-Salpeter equation. 
	Here the square  and the circle represent the potential $V$ 
	and the scattering amplitude $T$, respectively.}\label{pic:BSE}
	\end{figure}

In this section we collect the necessary formalism of the Bethe-Salpeter approach. We
 denote the in- and outgoing meson momenta by $q_1$ and $q_2$, respectively. Moreover
 the overall four-momentum is given by $p=q_1+p_1=q_2+p_2$, where $p_1$ and $p_2$ are
 the momenta of in- and out-going baryon, respectively. For the meson-baryon scattering
 amplitude $T(\slashed{q}_2, \slashed{q}_1; p)$ and chiral potential $V(\slashed{q}_2, \slashed{q}_1; p)$
 the integral equation to solve reads 
	\begin{align}\label{eqn:BSE}
	T(\slashed{q}_2, &\slashed{q}_1; p)= V(\slashed{q}_2, \slashed{q}_1; p) 
	+\nonumber\\
	&i\int\frac{d^d l}{(2\pi)^d}V(\slashed{q}_2, \slashed{l}; p) 
	S(\slashed{p}-\slashed{l})\Delta(l)T(\slashed{l}, \slashed{q}_1; p),
	\end{align}
 where $S$ and $\Delta$ represent the baryon (of mass $m$) and the meson (of mass $M$)
 propagator, respectively, and are given by $iS(\slashed{p}) ={i}/({\slashed{p}-m+i\epsilon})$
 and $i\Delta(k) ={i}/({k^2-M^2+i\epsilon})$. The BSE is depicted in Fig.~\ref{pic:BSE}.

So far we have suppressed the channel indices in the above 
 formulas, however since we are dealing with coupled channels, 
 $T$, $V$, $S$ and $\Delta$ are matrices in channel space (the propagators 
 are represented by diagonal matrices). In view of a later
 application to photoproduction off protons, we restrict ourselves to 
 meson-baryon channels with strangeness $S=0$ and electric charge $Q=+1\,$. 
 This leaves us with the following channels:
\begin{equation}
 p \pi^0,~n \pi^{+},~p\eta,~\Lambda K^+,~\Sigma^0 K^+,~\Sigma^+ K^0\, .
\end{equation}
Now let us specify the interaction kernel to be iterated by means of Eq.~\reff{eqn:BSE}.
 As explained above, we only include the contact-term contributions from
 $\Lagr^{(1)}_{\phi B}$ and $\Lagr^{(2)}_{\phi  B}$ and omit the Born terms. To
 our knowledge this is the first time these NLO corrections of the chiral potential
 are included and unitarized within the full relativistic BSE, without making use of
 the on-shell approximation or s-wave projection of the chiral potential,  so that
 also a p-wave is iterated. Separating the momentum space from channel space structures
 the chiral potential considered here takes the form:
	\begin{eqnarray}\label{eqn:coupling}
	V(\slashed{q}_2, \slashed{q}_1; p)&=&A_{WT}(\slashed{q_1}+\slashed{q_2})\nonumber\\
	&+&A_{14}(q_1\cdot q_2)+A_{57}[\slashed{q_1},\slashed{q_2}]+A_{M}(q_1\cdot q_2)\nonumber\\
	&+&A_{811}\big(\slashed{q_2}(q_1\cdot p)+\slashed{q_1}(q_2\cdot p)\big),
	\end{eqnarray}
 where the first matrix only depends on the meson decay constants $F_\pi ,\,F_K,\,F_{\eta}$,
 whereas  $A_{14}$, $A_{57}$, $A_{811}$ and $A_{M}$ also contain the NLO LECs as specified
 in \ref{app:coupling}. In going from the Lagrangian~\reff{eqn:LAGR} to the above vertex rule,
 we have left out some terms which are formally of third chiral order.

The loop diagrams appearing in the BSE Eq.~(\ref{eqn:BSE}) are in general
 divergent and require renormalization. In case of a strict chiral perturbation
 expansion, the terms can be renormalized  in a quite straightforward way,
 order by order, including at a given order of the calculation all the
 counterterms absorbing the loop divergencies. On the other hand the
 treatment of the divergencies of the BSE is known to be a complicated
 issue, see e.g.~\cite{Nieves:1999bx,Borasoy:2007ku}. Although
 the unitarization of the chiral potential provides us with large benefits
 regarding dynamically generated resonances, it relies on approximations of
 the kernel, which destroy some fundamental features of quantum field theory,
 such as crossing symmetry.

There are various ways to treat the divergent integrals and the large baryon
 mass scale appearing. Without going into details here, we preserve the
 analytic structure of the loop integrals by utilizing dimensional regularization
 and just replacing the divergent part by a subtraction constant.  
 The purely baryonic integrals are set to zero from the beginning. Thus, our treatment
 of the loop integrals is, in effect, similar to the EOMS regularization
 scheme advocated in \cite{Fuchs:2003qc}. As it was argued in \cite{Borasoy:2007ku}
 it is not possible to express the terms necessary to absorb the divergencies in
 the BSE as counterterms derived from a local Lagrangian. However it is possible
 to alter the loop integrals in the solution of the BSE in a way that is in
 principle equivalent to a proper modification of the chiral potential itself
 (for an explicit demonstration, see App.~F of \cite{PCB:Diss}). In this spirit
 we apply the usual $\overline{MS}$ subtraction scheme, keeping in mind that the
 modified loop integrals are still scale-dependent. This regularization scale
 ($\mu$) dependence would be canceled by the corresponding scale dependence of
 higher-order contact terms in the perturbative approach, but in our nonperturbative
 framework, the scale $\mu$ is used as a fitting parameter, reflecting the
 influence of higher order terms not included in our potential. Note that
 in \cite{Nieves:1999bx,Nieves:2001wt}, the  12 loop integrals
 (4 for each meson-baryon, meson and baryon case) appearing  there, gave rise to 12
 finite subtraction constants, which were then also  used as fitting parameters
 of their approach.

Having specified the kernel we are now ready to solve the Bethe-Salpeter equation.
 Given the structure of the kernel, its iteration via the BSE  induces the following
 form of the scattering amplitude,
	\begin{equation}
	T(\slashed{q}_2, \slashed{q}_1; p)=\sum_{i=1}^{20} \aleph_i \cdot \text{T}_i,
	\end{equation}
 where the coefficients T$_i$ are $6\times 6$ matrices in channel space,
 which only depend on the center-of-mass energy $\sqrt{s}$ after fixing
 the LECs, and $\aleph:=\Big(\slashed{q_1}${}, $\slashed{p}\slashed{q_1}${}, 
$\slashed{q_2}\slashed{p}\slashed{q_1}${}, $\slashed{q_2}\slashed{q_1}${}, 
$\slashed{p}\slashed{q_1}(q_2\cdot p)${}, $\slashed{q_1}(q_2\cdot{}p)${}, 
$\slashed{q_2}(q_1\cdot p)${}, $\slashed{q_2}\slashed{q_1}${}, 
$(q_1\cdot p)(q_2\cdot p)${}, $\slashed{p}(q_1\cdot~p)(q_2\cdot~p)$, 
$(q_1\cdot p)${}, $\slashed{p}(q_1\cdot p)${}, $(q_2\cdot q_1)${}, 
$\slashed{p}(q_2\cdot q_1)${}, $\slashed{q_2}\slashed{p}${},
$\slashed{q_2}${}, $\slashed{p}(q_2\cdot p)${}, $(q_2\cdot p)${}, $\mathds{1}${}, 
$\slashed{p}\Big)$
 is a vector in the $20$-dimensional space of invariant structures. Note that
 the scalar products are listed here as independent structures because we include
 the full off-shell dependence of the chiral potential in the BSE, which prevents
 us from writing them as simple functions of the Mandelstam variables $s$ and $t$.

On the other hand the above decomposition allows us to pull the coefficients
 T$_i$ out of the loop-integral in Eq. \reff{eqn:BSE}. Then these are fully
 determined by the solution of a linear system of equations in the
 space of invariant structures:
	\begin{equation}
	{\rm X}_{ij} {\rm T}_j = {\rm V}_i~, \qquad(i,j=1,\ldots,20)~,
	\end{equation}
 where the V$_i$ are coefficients of the chiral potential with respect to
 the invariant structures defined above and X is a $20\times 20$ matrix.
 The latter connects different structures of the space of invariant structures
 via loop integrations on the r.h.s.~of Eq. \reff{eqn:BSE}.
 Once the BSE has been solved, we can of course set the {\em external}\,
 four-momenta on their mass shells, leaving us with only two independent
 structures for the on-shell amplitude, i.e. $\mathds{1}$ and $\slashed{p}$.

		\section{Results and discussion}

Throughout the present work  we use the following numerical values (in GeV) for the
 masses and the meson decay constants:
 $F_\pi=F_\eta/1.3=0.0924$, $~F_K =0.113$, 
 $M_{\pi^0}=0.135$, $M_{\pi^+}=0.1396$,	
 $M_\eta=0.5478$, $M_{K^+}=0.4937$,
 $M_{K^0}=0.4977$, $m_p=0.9383$,
 $m_n=0.9396$, $m_\Lambda=1.1157$,
 $m_{\Sigma^0}=1.1926$ and $m_{\Sigma^+}=1.1894$.
 The baryon mass in the chiral limit, $m_0$ in Eq.~\reff{eqn:LAGR}, can be fixed
 to $1$~GeV without loss of generality, as any other value only amounts to a
 rescaling  of the unknown LECs.

There are 17 free parameters in the present approach, given by the 14 LECs,
 as well as three subtraction constants for the regularized loop integrals,
 corresponding to the logarithms of the undetermined regularization scales
 (in GeV), i.e.~$\log(\mu_\pi)$, $\log(\mu_K)$ and $\log(\mu_\eta)$. Here we
 take the regularization scale of each channel to be fixed by the respective
 meson, i.e.~in addition to $\mu_{\pi N}=:\mu_{\pi}$ and $\mu_{\eta N}=:\mu_{\eta}$,
 we take $\mu_{K\Sigma}=\mu_{K\Lambda} =: \mu_{K}$. The latter constraint appears to
 be natural in view of our forthcoming work on meson photoproduction, where
 loops are present in which a photon-induced $\Lambda\to\Sigma^{0}$ transition
 occurs.

For the fits, we consider  experimental data for s-wave $\pi N$ scattering up
 to $W=1.56$ GeV, i.e. partial wave amplitudes $S_{11}$ and $S_{31}$ (both real
 and imaginary parts) provided by the SAID--program at GWU, see \cite{Arndt:2006bf}.
 Comparing an earlier analysis by the Karlsruhe group \cite{Koch:1985bn}
 to the current
 one, we assign for the energies below $W=1.28$ GeV an absolute systematic error
 of $0.005$ and for higher energies an error of $0.030$ to the partial wave
 amplitudes. To some extent this is in agreement with error estimates done in
 \cite{Nieves:2001wt}, which are motivated by the expectation of pronounced
 three-body effects above the $\pi\pi N$ threshold.
	\begin{figure}[t]
	\begin{center}
	\includegraphics[width=0.95\linewidth]{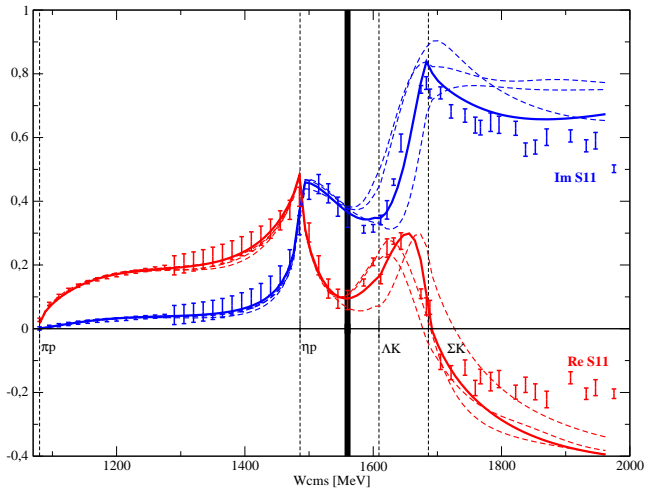}
	\end{center}
	\caption{Real and imaginary part of the $S_{11}$ partial wave amplitude
	 compared with the {SAID}-data (WI08-analysis). Full curves correspond to the best fit, the dashed ones to fits with slightly worse $\chi_{\rm dof}^2$. The bold vertical
	 line limits the region of the fit, where in the non-fit region single energy values are taken from the {SAID}-data.}\label{pic:S11}
	\end{figure}
For the best fit, found using the MINUIT library, with a $\chi_{\rm dof}^2=1.23$
 we obtain the following parameter set (all $b_i$ in GeV$^{-1}$)
	\begin{align}
	\log(\mu_\pi)&=+0.924,	&b_4&=-0.215,	&b_{10}&=+1.920,\nonumber\\
	\log(\mu_K)&=+0.581,  	&b_5&=-0.963,	&b_{11}&=-0.919,\nonumber\\
	\log(\mu_\eta)&=-0.218, &b_6&=+0.218,	&b_0&=-0.768,\nonumber\\
	b_1&=-0.082, 		&b_7&=-1.266,	&b_D&=+0.641,\nonumber\\
	b_2&=-0.118,  		&b_8&=+0.609,	&b_F&=-0.098,\nonumber\\
	b_3&=-1.890,		&b_9&=-0.633.	&\quad &
	\end{align}
 All parameters are of natural size and LECs agree with
 the estimates from the SU(3) to SU(2) matching relations provided in
 \cite{Mai:2009ce}. However we are only able to estimate the computational errors
 on the above parameters within the MIGRAD (MINUIT) minimization procedure,
 which appear to be negligible.

	\begin{figure}[t]
	\begin{center}
	\includegraphics[width=0.95\linewidth]{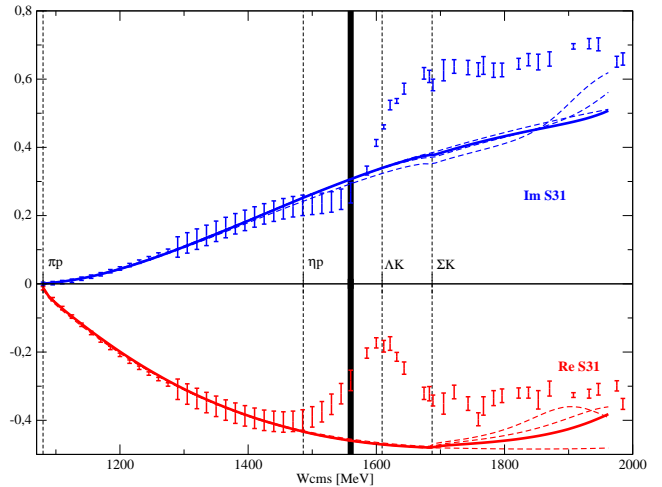}
	\end{center}
	\caption{Real and imaginary part of the $S_{31}$ partial wave amplitude
	 compared with the {SAID}-data (WI08-analysis). Full curves correspond to the best fit, the dashed ones to fits with slightly worse $\chi_{\rm dof}^2$. The bold vertical
	 line limits the region of the fit, where in the non-fit region single energy values are taken from the {SAID}-data.}\label{pic:S31}
	\end{figure}
In Figs.~\ref{pic:S11} and \ref{pic:S31} we present the result of our approach
 for  the $S_{11}$ and $S_{31}$ partial waves. As already seen in earlier
 publications on the BSE approach with leading order chiral potential
 \cite{Nieves:2001wt}, the low-energy region (e.g.~$\sqrt{s}<1.4$ GeV) is
 reproduced for both isospin $3/2$ and $1/2$ reasonably well. For the two
 s-wave scattering lengths, we obtain $a_{1/2} = 145.8 \times 10^{-3}/M_{\pi^+}$
 and $a_{3/2} = -91.6 \times 10^{-3}/M_{\pi^+}$, to  be compared with the
direct extraction of these scattering lengths from the GWU solution,
$a_{1/2} = (174.7\pm 2.2) \times 10^{-3}/M_{\pi^+}$ and 
$a_{3/2} = (-89.4\pm 1.7) \times 10^{-3}/M_{\pi^+}$.\footnote{We thank Ron Workman for
providing us with these values.}
The theoretically cleanest determination of these observables stems from the analysis 
 of pionic hydrogen and pionic deuterium
 data based on effective field theory~\cite{Baru:2010xn}, 
 $a_{1/2} = (179.9\pm 3.6) \times 10^{-3}/M_{\pi^+}$
 and $a_{3/2} = (-78.5\pm 3.2) \times 10^{-3}/M_{\pi^+}$.
The description of the $\pi N\,$ amplitude at low energies will certainly be improved by a more complete treatment of the Born terms, which is beyond the scope of this Letter. One might also think about constraining the well-known pion-nucleon scattering lengths, e.g. by adopting a matching procedure to the perturbative expansion. However, since we did not put a special weight on the threshold region in our fits, and the overall description of the partial waves seems to work well over a rather broad energy range, we regard the
 obtained results for the scattering lengths as satisfactory.

Moreover, and more importantly, within the fit region we reproduce
 the $S_{11}(1535)$, without any use of explicit vector meson resonances
 or even taking into account the $\pi\pi N$ channels as for example in
 \cite{Inoue:2001ip}. At the same time the $S_{31}(1620)$ resonance is
 not reproduced by our approach, which is in agreement with the current
 state of knowledge that the first $S_{31}$ resonance does not have a
 prominent dynamically generated component. To emphasize this we exclude
 the data on $S_{31}$ and recalculate the $\chi^2_{\rm dof}$ for the above
 parameter set, we end up with $\chi^2_{\rm dof}(S_{11})=0.59$.

At this point one realizes an even more interesting fact: After fixing the $S_{11}$
 partial wave in the energy region up to $\sqrt{s}=1.560$ GeV every curve with minimized
 $\chi^2_{\rm dof}$ possesses a second structure between $K\Lambda$ and $K\Sigma$
 threshold. Obviously this corresponds to the well-known $S_{11}(1650)$
 resonance and is predicted here only by demanding a good description
 in the low-energy and the first resonance region. To some extent
 this is in agreement with Ref.~\cite{Nieves:2001wt}, where the
 $S_{11}(1650)$ was reproduced in the fit of the phase shifts and
 inelasticities for the full region of $1.077 <\sqrt{s}/$GeV $<1.946$. While only
 the leading order chiral potential was considered there, the
 authors introduced additional parameters appearing for every loop
 integral. Apparently these parameters contain some of the information
 that has to be attributed to neglected terms of higher order in
 the chiral potential. Additionally, in contrast to our approach this
 method does not allow to identify the higher partial waves than the s-wave,
 which might become important for higher energies as emphasized in
 \cite{Borasoy:2007ku}.

	\begin{figure}[t]
	\begin{center}
	\includegraphics[width=0.99\linewidth]{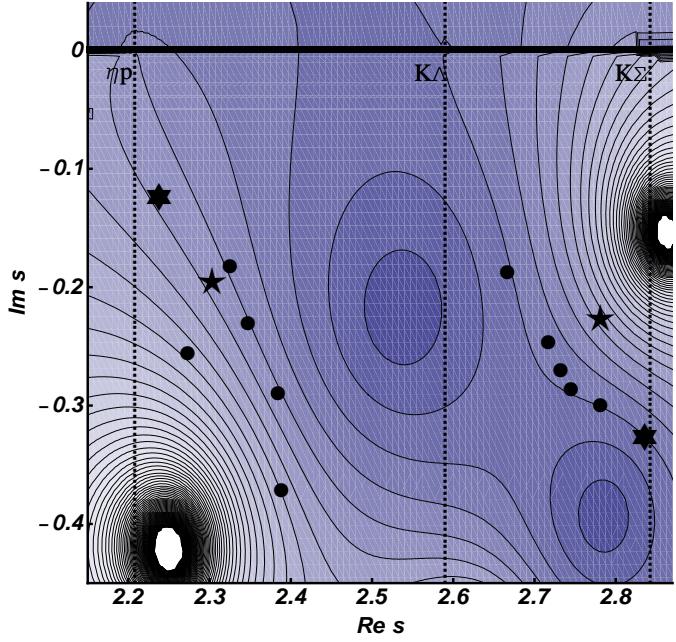}
	\end{center}
	\caption{(222-111) Riemann sheet of the $s$-plane. 
	 The five-star and the six-star correspond to the
	 values obtained in Ref.~\cite{Doring:2009yv} and
	 Ref.~\cite{Nieves:1999bx}, respectively, dots represent results
	 of phenomenological models listed in \cite{Yao:2006px}.}\label{pic:cntr1}
	\end{figure}
	\begin{figure}[t]
	\begin{center}
	\includegraphics[width=0.99\linewidth]{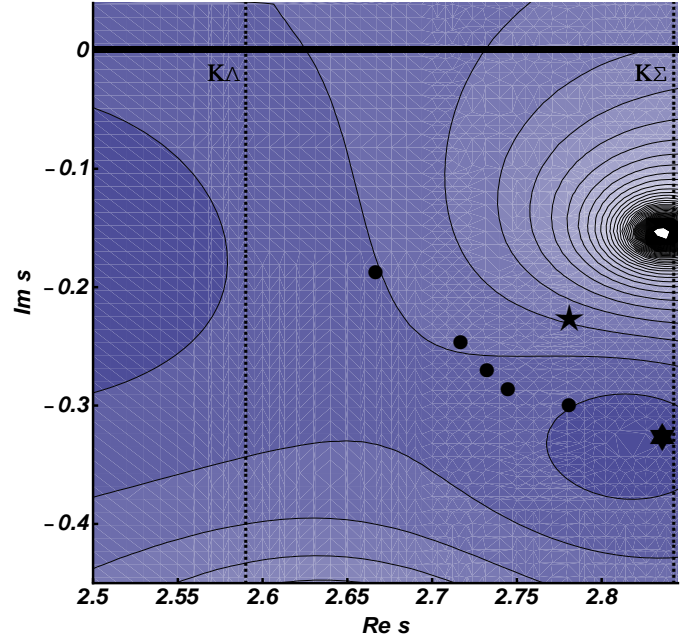}
	\end{center}
	\caption{(2222-11) Riemann sheet of the $s$-plane. 
	 The five-star and the six-star correspond to the
	 values obtained in Ref.~\cite{Doring:2009yv} and
	 Ref.~\cite{Nieves:1999bx}, respectively, dots represent results
	 of phenomenological models listed in \cite{Yao:2006px}.}\label{pic:cntr2}
	\end{figure}

In Figs.~\ref{pic:cntr1} and \ref{pic:cntr2} we present the modulus of
 the analytic   continuation of $T_{\pi N}^{11}$ into the complex $s$-plane. In
 Fig.~\ref{pic:cntr1} two poles appear on the (222-111) Riemann sheet,
 which labels the unphysical Riemann sheet connected to the physical (scattering) axis in the energy region
 between the third and fourth threshold, i.e. $(M_\eta+m_N)^2<s<(M_K+m_\Lambda)^2$.
 For the position of the two poles we extract:
	\begin{align}
	W_{1535}&= (1.506 -  0.140\text{\textit{ i}}) \text{ GeV}, \nonumber \\
	W_{1650}&= (1.692 - 0.046\text{\textit{ i}}) \text{ GeV}.
	\end{align}
 Choosing the (2222-11) Riemann sheet, i.e.~the unphysical sheet reached by analytic
 continuation from the region $(M_K+m_\Lambda)^2<s<(M_K+m_\Sigma)^2$, see
 Fig.~\ref{pic:cntr2}, we obtain one single pole structure,
 which is located at
	\begin{align}
		W_{1650}&= (1.682-0.042\text{\textit{ i}}) \text{ GeV}.
	\end{align}
 We conclude that the $S_{11}(1650)$ can also be described as a dynamically
 generated resonance, just like the $S_{11}(1535)$.
 
Clearly the uncertainty of our predictions grows with
 increasing energy. As a consequence of the sizeably increased
 computing time, when fitting the full amplitudes rather than the
 on-shell approximations to them, we are not able to perform a 
 full error analysis as e.g. done in Ref.~\cite{Borasoy:2006sr}   
 for $K^-p$ scattering. Still, we are able to get an indication of the
 error bands on the partial wave amplitudes. For this we present the second,
 third and fourth best fits in Figs.~\ref{pic:S11}
 and \ref{pic:S31} as dashed lines.
 However the error analysis deserves further
 studies. 

It is further interesting to analyze the structure of these states. To
 do that, we consider the on-shell scattering matrix in the vicinity of the
 two poles, where it takes the form
	\begin{equation}
	T_{ij}^{\rm on} (s) \simeq \frac{g_i g_j^{\ast}}{s- s_R}~,
	\end{equation}
 with $g_i ~(g_j)$ the complex coupling constant for the initial
 (final) transition of the meson-baryon system. For the $S_{11} (1535)$,
 we obtain the following ordering
	\begin{equation}\label{eq:S11}
	|g_{\Lambda K^+}|^2 > |g_{ p \eta}|^2 > |g_{\Sigma^+ K^0}|^2 \simeq 
	|g_{n \pi^+ }|^2 >   |g_{\Sigma^0 K^+}|^2 \simeq  |g_{ p \pi^0}|^2.
	\end{equation}
 We remark that the inequalities between couplings to different $\pi N$ and
 $K\Sigma$ channels are mostly due to Clebsch-Gordan coefficients
 in the associated isospin decompositions. However,
 isospin symmetry is not exact in the present approach. We find that
 the largest component is the $K\Lambda$ one and that the
 coupling to $\eta N$ is significantly bigger than the $\pi N$ ones,
 in agreement with the empirical fact that the $S_{11} (1535)$ couples
 dominantly to $\eta N$. The pattern for the $S_{11} (1650)$ looks different,
	\begin{equation}\label{eq:S31}
	|g_{\Sigma^+ K^0}|^2 > |g_{ p \eta}|^2 >  |g_{\Sigma^0 K^+}|^2 \simeq
	|g_{ n \pi^+}|^2 >   |g_{ p \pi^0}|^2 \gg |g_{\Lambda K^+}|^2,
	\end{equation}
 i.e.~for this resonance the $K\Sigma$ component is dominant and the
 $K\Lambda$ one is completely negligible, which for instance is indicated by the fact
 that the pole associated with the $S_{11}(1650)$ is accompanied by a second one on a neighboring sheet, with almost the same coordinates. As for the lower-lying resonance,
 the coupling to $N\eta$ is bigger than the one to $N\pi$.

		\section{Summary and outlook}

In this Letter, we have analyzed s-wave pion-nucleon scattering
in coupled-channel unitarized chiral perturbation theory. The
driving kernel includes all local interactions terms of first
and second order from the chiral effective Lagrangian. We consider
all two-body channels with strangeness zero and charge plus one, but do 
not include inelasticities generated from three-body $N\pi\pi$ states. 
The Bethe-Salpeter equation has been solved including the full 
off-shell dependence of the chiral potential. The parameters are
fitted to the real and imaginary part of the $S_{11}$ and the $S_{31}$
partial waves for cms energy below 1.56~GeV. We show that both the
$S_{11} (1535)$ and the $S_{11} (1650)$ are generated dynamically,
even though the fit range does only include the first resonance.
We have also analyzed the structure of these states, which
exhibit some marked differences
as indicated by the couplings given in
Eqs.~(\ref{eq:S11},\ref{eq:S31}). Quite differently, no resonance is 
generated in the $S_{31}$ partial wave. We consider this an
important step in our program of describing kaon photoproduction
from coupled-channel unitarized chiral perturbation theory.
Clearly, in the future more work is needed to properly include
the Born terms and to perform a systematic error analysis.

		\section*{Acknowledgments}
We thank M. D\"oring and B. Kubis for a careful reading of the manuscript.
One of the authors (M.M.) thanks B. Metsch, S. Kreuzer and H. van Pee
 for the great assistance with IT issues.
Partial financial support by the Helmholtz Association through funds provided
 to the virtual institute ``Spin and strong QCD'' (VH-VI-231),
 by the European Community-Research Infrastructure Integrating Activity 
 ``Study of Strongly Interacting Matter''
 (acronym HadronPhysics2, Grant Agreement n.~227431) under the Seventh 
 Framework Programme of the EU,
 and by DFG (SFB/TR 16, ``Subnuclear Structure of Matter'') is gratefully
 acknowledged.

\appendix
\section{Couplings}
\label{app:coupling}
For the channel indices $\{b,j;i,a\}$ corresponding to the process $\phi_iB_a\rightarrow\phi_jB_b$ the relevant coupling matrices read
\begin{align*}
&A_{WT}^{b,j;i,a}=-\frac{1}{4F_j F_i}\langle\lambda^{b\dagger}[[\lambda^{j\dagger},\lambda^{i}],\lambda^{a}]\rangle,\\
&A_{14}^{b,j;i,a}=-\frac{2}{F_j F_i}\Big(~ b_1\Big(\langle\lambda^{b\dagger}[\lambda^{j\dagger},[\lambda^{i},\lambda^{a}]]\rangle 
+\langle\lambda^{b\dagger}[\lambda^{i},[\lambda^{j\dagger},\lambda^{a}]]\rangle\Big)\\
& + b_2\Big(\langle\lambda^{b\dagger}\{\lambda^{j\dagger},[\lambda^{i},\lambda^{a}]\}\rangle 
+\langle\lambda^{b\dagger}\{\lambda^{i},[\lambda^{j\dagger},\lambda^{a}]\}\rangle\Big)\\
&+b_3\Big(\langle\lambda^{b\dagger}\{\lambda^{j\dagger},\{\lambda^{i},\lambda^{a}\}\}\rangle 
+\langle\lambda^{b\dagger}\{\lambda^{i},\{\lambda^{j\dagger},\lambda^{a}\}\}\rangle\Big)+
2b_4 \langle\lambda^{b\dagger}\lambda^{a}\rangle \langle\lambda^{j\dagger}\lambda^{i}\rangle
\Big),\\
&A_{57}^{b,j;i,a}=-\frac{2}{F_j F_i}\Big(~
b_5\langle\lambda^{b\dagger}[[\lambda^{j\dagger},\lambda^{i}],\lambda^{a}]\rangle+
b_6\langle\lambda^{b\dagger}\{[\lambda^{j\dagger},\lambda^{i}],\lambda^{a}\}\rangle\\
&+
b_7\Big(\langle\lambda^{b\dagger}\lambda^{j\dagger}\rangle \langle\lambda^{i}\lambda^{a}\rangle+
\langle\lambda^{b\dagger}\lambda^{i}\rangle \langle\lambda^{a}\lambda^{j\dagger}\rangle\Big)
\Big),\\
&A_{811}^{b,j;i,a}=-\frac{1}{F_j F_i}\Big(~
b_8\Big(
\langle\lambda^{b\dagger}[\lambda^{j\dagger},[\lambda^{i},\lambda^{a}]]\rangle 
+\langle\lambda^{b\dagger}[\lambda^{i},[\lambda^{j\dagger},\lambda^{a}]]\rangle
\Big)  \\ 
&+b_9\Big(
\langle\lambda^{b\dagger}[\lambda^{j\dagger},\{\lambda^{i},\lambda^{a}\}]\rangle 
+\langle\lambda^{b\dagger}[\lambda^{i},\{\lambda^{j\dagger},\lambda^{a}\}]\rangle\Big)\\
& +b_{10}\Big(\langle\lambda^{b\dagger}\{\lambda^{j\dagger},\{\lambda^{i},\lambda^{a}\}\}\rangle 
+\langle\lambda^{b\dagger}\{\lambda^{i},\{\lambda^{j\dagger},\lambda^{a}\}\}\rangle\Big)\\
& +
2b_{11}\langle\lambda^{b\dagger}\lambda^{a}\rangle \langle\lambda^{j\dagger}\lambda^{i}\rangle
\Big),\\
&A_{M}^{b,j;i,a}=-\frac{1}{2 F_j F_i}\Big(~
b_D\Big(
\langle\lambda^{b\dagger}\{\{\lambda^{j\dagger},\{\bar{\mathcal{M}},\lambda^{i}\}\},\lambda^{a}\}\rangle
\\ &\qquad\qquad\qquad\qquad +
\langle\lambda^{b\dagger}\{\{\lambda^{i},\{\bar{\mathcal{M}},\lambda^{j\dagger}\}\},\lambda^{a}\}\rangle
\Big)\\
& +b_F\Big(
\langle\lambda^{b\dagger}[\{\lambda^{j\dagger},\{\bar{\mathcal{M}},\lambda^{i}\}\},\lambda^{a}]\rangle+
\langle\lambda^{b\dagger}[\{\lambda^{i},\{\bar{\mathcal{M}},\lambda^{j\dagger}\}\},\lambda^{a}]\rangle
\Big)\\
& +2b_0\Big(
\langle\lambda^{b\dagger}\lambda^{a}\rangle \langle[\lambda^{j\dagger}\lambda^{i}]\bar{\mathcal{M}}\rangle
\Big)
\Big),
\end{align*}
where $\lambda$ denote the $3\times 3$ channel matrices (e.g. $\phi =
\phi^{i}\lambda^{i}$ for the physical meson fields), the $F_i$ are the decay constants of the meson in the respective channel, and $\langle\ldots\rangle$
denotes the trace in flavor space. 
Moreover, $\bar{\mathcal{M}}$ is obtained from the quark mass matrix $\mathcal{M}$
via the Gell--Mann Oakes Renner 
relations, and given in terms of the meson masses as follows,
$\bar{\mathcal{M}}=\frac{1}{2}{\rm diag}(M_{K^+}^2 - M_{K^0}^2 + M_{\pi^0}^2,
M_{K^0}^2 - M_{K^+}^2 + M_{\pi^0}^2, M_{K^+}^2 + M_{K^0}^2 - M_{\pi^0}^2)\,$.


\end{document}